\documentstyle[11pt]{article}
\newcommand{\bmat}{\left(\begin{array}}
\newcommand{\emat}{\end{array}\right)}
\newcommand{\be}{\begin{equation}}
\newcommand{\ee}{\end{equation}}
\newcommand{\ba}{\begin{array}}
\newcommand{\ea}{\end{array}}
\newtheorem{dfn}{Definition}[section]
\newtheorem{lemma}{Lemma}[section]

\newtheorem{theorem}{Theorem}[section]
\author{I G Korepanov\\
\small Chelyabinsk 454080, pr.~Lenina 78-A kv.~45, Russia}
\date{}
\title{A Dynamical System Connected with Inhomogeneous 6-Vertex Model}
\begin{document}

\maketitle
\begin{abstract}
A completely integrable dynamical system in discrete time is studied by
means of algebraic geometry. The system is associated with factorization
of a linear operator acting in a direct sum of three linear spaces into a
product of three operators, each acting nontrivially only in a direct sum
of two spaces, and the following reversing of the order of factors. There
exists a reduction of the system interpreted as a classical field theory
in $2+1$-dimensional space-time, the integrals of motion coinciding, in
essence, with the statistical sum of an inhomogeneous 6-vertex free-fermion
model on the 2-dimensional kagome lattice (here the statistical sum is a
function of two parameters). Thus, a connection with the ``local'', or
``generalized'', quantum Yang---Baxter equation is revealed.
\end{abstract}

This paper is devoted to an exactly solvable model in discrete time
associated with factorization of a certain matrix in a product of {\em
three\/} matrices and the following change of the order of those matrices
into the inverse one, and also to a reduced case of this model, when it
becomes a sort of field theory in $2+1$-dimensional wholly discrete
space-time. The generating function of the motion integrals in this reduced
case is nothing else but a statistical sum of an inhomogeneous 6-vertex
model (depending on 2 parameters) on the kagome lattice. One can say thus
that the reduced dynamical system describes an evolution in discrete time
of the 6-vertex model Boltzmann weights which conserves the statistical sum.
The evolution is local, which means that the value of each weight in the
moment~$\tau+1$ is affected only by (itself and) several neighboring weights
in the moment~$\tau$. The weights satisfy the ``free fermion'' condition
known from the quantum Yang---Baxter equation (QYBE) theory.

The evolution of weights can be seen as a certain realization of ideas of
the paper~\cite{MN} concerning the ``localization'' of such equations as
QYBE. As is known, the usual QYBE
$$R_{12}L_{13}M_{23}=M_{23}L_{13}R_{12}$$
contains the same operators in the L.H.S. as in the R.H.S. (the indices
pointing to different linear spaces in whose tensor product an operator
acts). The local QYBE
\be R_{12}L_{13}M_{23}=M'_{23}L'_{13}R'_{12} \label{1lqyb} \ee
has different operators in each side. Note, however, that the fundamental
equation in the present work is an equation differing from~(\ref{1lqyb})
in that the operators in it act not in a tensor product, but in a {\em
direct sum\/} of three spaces! This will be seen in Section~\ref{1secdef},
where we will have to represent as a product of three matrices some matrix
obtained at the previous step of evolution as a similar product, but in the
inverse order. How a tensor product arises from a direct sum, can be
explained by using the Clifford algebras. However, in this paper it is not
done, and the 6-vertex model is obtained through some consideration of
topological character.

Note also that the connection between the statistical physics models of
free-fermion type and completely integrable models is known for a long time
now---see~\cite{Jimbo-et-al} and the works cited therein. The first of the
author's works on this subject was~\cite{Korepanov-dimers}, where a
statistical sum conserving evolution of the weights was proposed for a flat
dimer model. In both papers, \cite{Korepanov-dimers} and the present one,
the tool of investigation is algebro-geometrical objects: algebraic curves
and divisors. These objects are well known in mathematical physics since the
works on finite-gap integration of soliton equations appeared (see e.g.\
the book~\cite{NMPZ}). In dealing with them, I try to hold to a reasonable
degree of strictness, in order not to overload the text with unnecessary
details.

\section{Definition of dynamical system} \label{1secdef}

Let
\be{\cal A}=\bmat{ccc}A&B&C\\D&F&G\\H&J&K \emat \label{101} \ee
be a block matrix acting in the linear space of $m+n+r$--dimensional
complex column vectors, so that, e.g., $A,F$ and $K$ are square matrices
of sizes $m \times m,\; n \times n,\; r \times r$ respectively.
Consider the problem of factorization of matrix $\cal A$ into a
product of the form
\be {\cal A} = {\cal A}_1 {\cal A}_2 {\cal A}_3, \label{102} \ee
where
\be \ba {c} {\cal A}_1 = \bmat {ccc}A_1&B_1&{\bf 0}\\C_1&D_1&{\bf 0} \\
{\bf 0}&{\bf 0}&{\bf 1} \emat, \quad
{\cal A}_2= \bmat {ccc} A_2&{\bf 0}&B_2\\{\bf 0}&{\bf 1}&{\bf 0}\\
C_2&{\bf 0}&D_2 \emat, \medskip   \\
{\cal A}_3=\bmat {ccc}{\bf 1}&{\bf 0}&{\bf 0}\\
{\bf 0}&A_3&B_3\\{\bf 0}&C_3&D_3 \emat. \ea \label{103} \ee
The factorization~(\ref{103}) may be seen as a generalization of the
factorization of an  orthogonal rotation in the 3-dimensional space
into rotations through the ``Euler angles''. However, no orthogonality
conditions will be considered in this paper.

One can obtain from the factorization~(\ref{102}) other factorizations
of the same kind by using the following transformation of the triple
${\cal A}_1,{\cal A}_2,{\cal A}_3$:
\be \ba {c}{\cal A}_1 \rightarrow {\cal A}_1 \bmat {ccc} M_1&{\bf 0}&{\bf 0}\\
{\bf 0}&M_2&{\bf 0}\\
{\bf 0}&{\bf 0}&{\bf 1} \emat , \quad {\cal A}_3 \rightarrow  \bmat {ccc}
{\bf 1}&{\bf 0}&{\bf 0}\\{\bf 0}&{M_2}^{-1}&{\bf 0}\\
{\bf 0}&{\bf 0}&{M_3}^{-1} \emat {\cal A}_3,  \medskip \\
{\cal A}_2 \rightarrow \bmat {ccc} {M_1}^{-1}&{\bf 0}&{\bf 0}\\
{\bf 0}&{\bf 1}&{\bf 0}\\{\bf 0}&{\bf 0}&{\bf 1} \emat
{\cal A}_2
\bmat {ccc}{\bf 1}&{\bf 0}&{\bf 0}\\{\bf 0}&{\bf 1}&{\bf 0}\\
{\bf 0}&{\bf 0}&M_3 \emat,  \ea \label{104} \ee
$ M_1, M_2$ and $M_3$ being arbitrary non-degenerate matrices of proper
sizes.

\begin{lemma} \label{1lem1} { For a generic matrix $\cal A$
factorization~(\ref{102}) is (if exists) unique to within the
transformations~(\ref{104}).  }
\end{lemma}

{\it Proof}. First, let us specify that matrix $\cal A$ will be called
generic with respect to this lemma if it a) is non-degenerate and
b) {\em allows}\/ a factorization~(\ref{102}) with matrices ${\cal A}_2$
and ${\cal D}_2$ non-degenerate. Let
\be {\cal A}= {\cal A}'_1{\cal A}'_2{\cal A}'_3\label{105} \ee
be another factorization of the same kind.
 From~(\ref{102}) and (\ref{105}) one finds
\be {\cal A}'_2= {\cal A}''_1 {\cal A}_2 {\cal A}''_3, \label{106} \ee
where
\be {\cal A}''_1= {({\cal A}'_1)}^{-1} {\cal A}_1, \;\;
{\cal A}''_3=  {\cal A}_3 {({\cal A}'_3)}^{-1}. \label{107}\ee
Let us denote the blocks in the dashed matrices by the same letters
$ A_1, \ldots,$ $D_3 $ as in equalities~(\ref{103}) with proper number of
dashes added to them. The relation~(\ref{106}) is rewritten as
\be
\bmat{ccc} A'_2&{\bf 0}&B'_2\\{\bf 0}&{\bf 1}&{\bf 0}\\C'_2&{\bf 0}&D'_2
\emat=
\bmat{ccc} A''_1 A_2 & A''_1B_2C''_3+B''_1A''_3 & A''_1B_2D''_3+B''_1B''_3\\
C''_1A_2 & C''_1B_2C''_3+D''_1A''_3 & C''_1B_2D''_3+D''_1B''_3\\
C_2 & D_2C''_3 & D_2D''_3 \emat  .
\label{108} \ee
 From here, one obtains at once the equalities
$$C'_2=C_2,\; C'_1={\bf 0},\; C''_3={\bf 0}.$$
Taking this into account, one finds from the block in 2nd row and 2nd
column that
$$D''_1A''_3={\bf 1}.$$
Thus, $D''_1$ and $A''_3$ are non-degenerate. Now the blocks just above
the main diagonal yield
$$ B''_1={\bf 0},\; B''_3={\bf 0}. $$
So, the matrices ${\cal A}''_1$ and ${\cal A}''_3$~(\ref{107}) are
block--diagonal. In is easy to see that this means exactly that
${\cal A}'_1, {\cal A}'_2, {\cal A}'_3$ are obtained from
${\cal A}_1, {\cal A}_2, {\cal A}_3$ by the transformation~(\ref{104}).
The lemma is proved.

Now let us construct, starting from the block martix ${\cal A}$,
new matrix ${\cal B}$  by following means: factorize  ${\cal A}$
into the product~(\ref{102}) and set
\be {\cal B}={\cal A}_3{\cal A}_2{\cal A}_1.  \label{109} \ee
 From the above considerations it is seen that the matrix ${\cal B}$
is determined to within the transformations
\be {\cal B}\rightarrow \bmat{ccc}{M_1}^{-1}&{\bf 0}&{\bf 0}\\
{\bf 0}&{M_2}^{-1}&{\bf 0}\\
{\bf 0}&{\bf 0}&{M_3}^{-1} \emat {\cal B}
\bmat {ccc} M_1&{\bf 0}&{\bf 0}\\ {\bf 0}&M_2&{\bf 0}\\
{\bf 0}&{\bf 0}&M_3 \emat.\label{110} \ee
Let us call such transformations, as applied to the block matrices
here, the gauge transformations. The following simple but important
observation is valid: {\em if matrix ${\cal A}$ itself undergoes a gauge
transformation, this in no way affects the set of matrices ${\cal B}$
obtained from formula~(\ref{109})}.

It will be shown in Section~\ref{1secevd} that factorization~(\ref{102})
does exist for a generic matrix {\cal A}. This factorization will be
constructed by means of algebraic geometry. Taking this into account,
we are ready now to define the dynamic system
that we are going to examine. Let ${\cal M}$ be the set of block
matrices~(\ref{101}) taken to within gauge transformations~(\ref{110}),
or, using stricter language, the set of equivalence classes of such
matrices with respect to transformations~(\ref{110}). The set ${\cal M}$
will be our ``phase space''. Then, the birational mapping $f$ is defined
on the set ${\cal M}$ that brings into correspondence with a matrix
${\cal A}$, factorized into the product~(\ref{103}),
the matrix ${\cal B}$ factorized into the product~(\ref{109}).
Let us now bring into consideration the ``discrete time'' $\tau$
taking integer values and say that to the trasition from time $\tau$ to
time $\tau+k$ corresponds the mapping
$$\underbrace{f\circ \ldots f}_{k\ \rm times}.$$

\section{Invariant algebraic curve of matrix $\cal A$ and some divisors
on it}
\label{1seccrv}

The dynamical system of Section~\ref{1secdef} turns out to be
completely integrable.
To be exact, {\em an invariant curve}  $\Gamma $ can be constructed out
of matrix $\cal A$, together with a divisor  $\cal D$ on it.
In terms of these algebro-geometrical objects, the evolution is as follows:
$\Gamma$ does not change, while  $\cal D$---more precisely, its linear
equivalence class---depends linearly on the discrete time $\tau$.

Let us start from the definition of the curve $\Gamma$. The word
``invariant'' in this definition will be justified in Section~\ref{1seccrv}.

\begin{dfn} \label{1dfn1} The invariant curve $\Gamma$ of the operator
$\cal A$ of the form~(\ref{101}) is an algebraic curve in
${\bf C}P^1\times{\bf C}P^1\times{\bf C}P^1$ (i.e.\ in the space of
three complex variables $u,v,w$, each allowed also to take value $\infty$)
given by equations
\be \det({\cal A}-\bmat{ccc} u{\bf 1}_m &  {\bf 0} & {\bf 0}\\
{\bf 0} & v{\bf 1}_n & {\bf 0}\\ {\bf 0} & {\bf 0} & w{\bf 1}_r \emat )=0,
\label{111} \ee
\be v=uw. \label{112} \ee
\end{dfn}

Here a subscript of each {\bf 1} means the size of corresponding unity
martix, while {\bf 0} denotes rectangular zero matrices of
different sizes. Strictly speaking, equations (\ref{111}, \ref{112})
define the ``finite part'' of the curve $\Gamma$, the whole curve $\Gamma$
being its closure in Zariski topology.

The equality~(\ref{111}) means that a column vector
${\cal X}= \bmat{c} X\\Y\\Z \emat$ exists, with $X,Y,Z$ column vectors
of dimensions $m,n$ and $r$ correspondingly, such that
\be {\cal A} \bmat{c}X\\Y\\Z \emat = \bmat{c}uX \\ vY \\ wZ \emat.
\label{113} \ee
Such vectors  $\cal X$ form a one-dimensional holomorfic bundle over
$\Gamma$.

The next lemma shows the structure of the zero and pole divisors of
functions $u,v,w$. For these divisors, the notations $(u)_0,(u)_{\infty}$
etc.\ are used.

\begin{lemma}\label{1lem2} {There exist such effective divisors (i.e.\
finite sets of points)  ${\cal D}_1, \ldots , {\cal D}_6$ in the curve
$\Gamma$ that
\be \ba{lll} (u)_{\infty}= {\cal D}_1+{\cal D}_2,&
(v)_{\infty}= {\cal D}_1+{\cal D}_3,& (w)_{\infty}= {\cal D}_3+{\cal D}_4,\\
(u)_0= {\cal D}_4+{\cal D}_6,& (v)_0= {\cal D}_5+{\cal D}_6, &
(w)_0= {\cal D}_5+{\cal D}_2. \ea  \label{114} \ee
${\cal D}_3$ and ${\cal D}_5$ are of degree   $m, \quad {\cal D}_2$ and
${\cal D}_4$ are of degree $n, \quad {\cal D}_1$ and ${\cal D}_6$
are of degree $r$. Generally, {\em all} points included in divisors
${\cal D}_1, \ldots , {\cal D}_6$ are different from each oter.
}\end{lemma}

{\it Proof}. Consider, e.g., the case $u=0, w \not= 0,
w \not= \infty$. Then, according to~(\ref{112}), $v=0$.
The equality~(\ref{113}) turns into the following system:
$$ \left\{ \ba{l} AX+BY+CZ=0, \\ DX+FY+GZ=0, \\ HX+JY+KZ=wZ. \ea \right.$$

One can express $X$ and  $Y$ through $Z$ (e.g.,
$Y=-(F-DA^{-1}B)^{-1}(G-DA^{-1}C)Z)$
and then substitute these expressions into the third one.One will come to
an equation of the form
\be \tilde K Z=wZ \label{115} \ee
which has $r$ characretistic roots  $w_1, \ldots ,w_r$, different from
each other in general case. This is how  $r$ points $(0,0,w_1),
\ldots ,(0,0,w_r)$ of divisor ${\cal D}_6$ are obtained. The other
divisors in~(\ref{114}) arise in a similar way. The lemma
is proved.

The vector $\bmat{c}X\\Y\\Z \emat $ in (\ref{113}) is determined
up to a scalar factor which may depend on the point in the curve
$\Gamma$.  So, this vector  can be normalized by setting its first
coordinate identically equal to unity~(cf. \cite{Krichever}).
$\bmat{c} X\\Y\\Z \emat$ becomes then a meromorphic vector on $\Gamma$
with a certain pole divisor $\cal D$. However,  $X,Y$ and $Z$ taken
separetely satisfy stronger restrictions, as the following lemma shows.
In the lemma, $(f)$ denotes the divisor of a function $f$ (zeros enter
with the $+$ sign, poles with the $-$ sign, as usual).

\begin{lemma} \label{1lem3} The column vector $X$ consists of
functions $f$ such that
\be (f)+{\cal D}-(u)_{\infty}\geq0; \label{116} \ee
the column vector $Y$ consists of functions $f$ such that
\be (f)+{\cal D}-(v)_{\infty}\geq0; \label{117} \ee
the column vector $Z$ consisits of functions $f$ such that
\be (f)+{\cal D}-(w)_{\infty}\geq0. \label{118} \ee
\end{lemma}
{\it Proof}. One can see immediately from the formula~(\ref{113}) that
the vector  $uX$  entering into R.H.S. cannot grow faster than the vector
$\bmat{c} X\\Y\\Z \emat$ in L.H.S. in such points where $u=\infty$.
This is exactly what the inequality~(\ref{116}) states.
The inequalities~(\ref{117}) and (\ref{118}) are proved similarly.

Now the time has come to make it sure that the curve $\Gamma$,
for a generic matrix $\cal A$, is a smooth irreducible curve.
One may wish also to calculate its genus in some simple way.
To do that, we are now going to examine a relatively simple
particular case of the matrix $\cal A$, although at the time ``generic''
enough to make sure that such its features as genus and the degree of
divisors are the same for matrices in some Zariski neighborhood.

Thus, let all the matrix elements of $\cal A$ equal zero except the ones
lying, first, in the main diagonal and, second, in the ``broken''
diagonal parallel to the main one (for these latter matrix elements,
the difference between the numbers of a column and a row must be
some constant modulo $m+n+r$). The elements in the main diagonal will
be denoted as $a_1, \ldots a_m, f_1, \ldots f_n,$ $k_1, \ldots k_r$;
and let the elements in the broken diagonal be all equal to the
same complex number $s$:
\be {\cal A} = \left(
{\linethickness{0.05pt}
\ba{ccc|ccc|ccc}
a_1 & & & & s & & & & \\
&\ddots &&&& \ddots &&&\\
&&a_m &&&& \ddots && \\ \hline
&&&f_1&&&& \ddots& \\
&&&& \ddots &&&& s \\
s&&&&&f_n&&&\\ \hline
& \ddots &&&&& k_1 && \\
&& \ddots &&&&& \ddots & \\
&&& s &&&&&k_r \ea
}
\right). \label{119} \ee
It does not matter through which blocks exactly the ``broken'' diagonal
passes.

For the finite  $u,w$, the curve $\Gamma$ now examined is given by
equation (resulting from the substitution of~(\ref{112}) into (\ref{111}))
\be F(uw) \equiv \prod^m_{\alpha =1}(a_{\alpha}-u) \cdot
\prod^n_{\beta =1}(f_{\beta}-uw)\cdot
\prod^r_{\gamma =1}(k_{\gamma}-w)\pm s^{m+n+r}=0. \label{120} \ee
As is known, in singular points
\be \left\{ \ba{l} \partial F / \partial u=0, \quad \\
\partial F / \partial w=0. \ea \right. \label{121} \ee
The system~(\ref{121}) has a finite number of solutions, and, changing $s$
in~(\ref{120}), one can make these solutions not to lie in the curve
$\Gamma$, which thus will be free of singularities for finite $u,v$.
It is an easy exercise to show that there are no singullarities when $u$
or $w$ is infinite as well.

Returning now to general martices $\cal A$  and curves $\Gamma$, let us
note that it cannot be that the system~(\ref{121}) or its substitute
in the neighborhood of infinite $u$ or $w$ possesses solutions
{\em in the curve}\/ $\Gamma$ in general case and does not possess them
in a particular case. Thus, the smoothness of $\Gamma$ for a generic
$\cal A$ is clear. As for irreducibility, to prove it let us examine the
natural projection of $\Gamma$ onto the Rimann sphere ${\bf C}P^1$
of the variable $u$. This projection is an $n+r$--sheet cover,
and if $\Gamma$ consisted of two or more components, the sheets of the cover
would split into groups belonging to each component.
To prove that it is not so in the general case, it is enough to
present an example where it is not so. To do this, take  $\cal A$ of the
form~(\ref{119}) and, moreover, put $f_1=\ldots =f_n=k_1=\ldots =k_r=0$.
Equation~(\ref{120}) then becomes
$$ w^{n+r}u^n \prod^m_{\alpha =1} (a_{\alpha}-u) \pm s^{m+n+r}=0.$$
Let  $a_1 \not= 0$ and not coinside with other $a_{\alpha}$.
Then in a neighborhood of the point $(u,w)=(a_1,\infty) \quad w$ behaves
like
$$ w^{-1}\sim (a_1-u)^{1/(n+r)}.$$
 From here one sees that all the mentioned $n+r$ sheets belong to a
{\em single} component, i.e.\ the irreducibility of $\Gamma$ is proved.

Now let us denote the number of branch points of the cover
$\Gamma \rightarrow {\bf C} \ni u$ \/ as  $b$. Then  the {\bf genus}
of the curve \/ $\Gamma$, according to the Riemann---Hurwitz formula, is
\be g=1-n-r+\frac{b}{2}. \label{122} \ee
Our next aim is to express  $b$  and  $g$ through $m,n$ and $r$.

\begin{lemma}\label{1lem4} {The degree of the vector
$\bmat{c}X\\Y\\Z \emat $ pole divisor $\cal D$ is $m+b/2$.
}
\end{lemma}

{\it Proof}. Write out the equation~(\ref{113}) ``explicitly'':
\be \left. \ba{c}  AX+BY+CZ=uX,\\ DX+FY+GZ=vY,\\
HX+JY+KZ=wZ. \ea \right\} \label{123} \ee
Expressing  $X$ through $Y$ and $Z$ by means of the first of these
equations and substituting into the rest, one finds:
\be \bmat{cc}\! u^{-1} \left( D(u-A)^{-1}B+F\right) & u^{-1}
\left( D(u-A)^{-1}C+G\right)\! \\
\!H(u-A)^{-1}B+J & H(u-A)^{-1}C+K\! \emat\! \bmat{c}\!Y\!\\\!Z\! \emat\! =
\!w\! \bmat{c}\! Y\!\\\!Z\! \emat\!. \label{124} \ee
To a generic $u$ correspond  $n+r$ different $w=w_1, \ldots w_{n+r}$,
and the corresponding $n+r$ vectors $\bmat{c} Y\\Z \emat $
are linearly independent as eigenvectors of the matrix in L.H.S.
of~(\ref{124}). An easy check shows that in the ``suspicious'', from
the standpoint of equation~(\ref{124}), points  $u=0, \infty$, and such
points where  $\det (u-A)=0$, there exist  $n+r$ linearly independent
vectors $\bmat{c} Y\\Z \emat$ as well.

Consider a determinant
$$ d= \left|\ba{ccc} Y(w_1) & \ldots & Y(w_{n+r})\\
Z(w_1) & \ldots & Z(w_{n+r}) \ea \right|. $$
Given $u$, it changes its sign under odd permutations of $w$'s.
This means that $d\/^2$ is a function of $u$ only. From the above one sees
that {\em  the number of zeros of function $d\/^2(u)$  equals $b$}, because
it is in the branch points and only in them that $d\/^2(u)$  vanishes. Thus,
the number of function $d^2(u)$ poles also equals $b$, and the pole divisor
degree of the meromorphic vector $\bmat {c} Y\\Z \emat $  is $b/2$. All this
consideration is perfectly standard, see~\cite{Krichever,Krichever77}.

Finally, from the 2nd and 3rd equations of the system~(\ref{123}) one sees
that the vector $X$
has its poles, when  $Y$ and $Z$ are finite, in those points of
$\Gamma$  where $v=w=\infty $ and only in them. Formulae~(\ref{114})
show that these are the points of divisor ${\cal D}_3$, and there are
$m$ of them. This adds  $m$ more poles to $b/2$ already present, and
with this the proof of the lemma is complete.

Let us turn again to matrices  $\cal A$ of the form~(\ref{119}). For such
matrix, it is easy to find the vector $\bmat{c}X \\ Y \\Z \emat=\cal X$  in
a given point $(u,v,w) \in \Gamma$. Let us assume that the ``broken''
diagonal is the one adjacent to the main diagonal, so that there is only
one letter $s$ in the lower left corner. Then the following holds
for the vector $\cal X$ coordinates:
$$ \ba{l}
(a_1-u)X_1+sX_2=0,\\
(a_2-u)X_2+sX_3=0,\\ \dotfill \\
(a_m-u)X_m+sY_1=0,\\
(f_1-v)Y_1+sY_2=0,\\ \dotfill \\
(f_n-v)Y_n+sZ_1=0, \\
(k_1-w)Z_1+sZ_2=0, \\ \dotfill \\
(k_n-w)Z_n+sX_1=0. \ea $$

 From here the ratios between vector $\cal X$ coordinates are readily seen.
Assuming the normalization condition $X_1\equiv 1$, one finds out that
$\cal X$ has the poles a) of the order $m$ in $n$
points $(u,v,w)=(\infty, f_{\beta}, 0)$ and b) of the order $m+n$
in $r$ points $(u,v,w)=(\infty, \infty, k_{\gamma})$. In all, $\cal X$
possesses thus $mn+mr+nr$ poles, taking their multiplicities into account.
Recalling Lemma~\ref{1lem4} and formula~(\ref{122}), one can now find
the genus $g$
of the curve as well. As a matrix $\cal A$ of the form~(\ref{119}) is
``generic enough'', the results on the degree of divisor $\cal D$ of the
vector $\cal X$ and genus $g$ of the curve apply also to curves corresponding
to generic matrices $\cal A$. Let us formulate them as the following lemma.

\begin{lemma} \label{1lem5}
For a generic matrix $\cal A$, the genus of the curve  $\Gamma$ is
\be g=mn+mr+nr-m-n-r+1, \label{125} \ee
while the degree of divisor $\cal D$ of the meromorphic vector
$\cal X=\bmat{c}X \\ Y \\Z \emat$  is
\be mn+mr+nr=g+m+n+r-1.\label{126} \ee
\end{lemma}

Thus, in this section we have constructed, for a given matrix $\cal A$,
an algebraic curve $\Gamma$ and a bundle of vectors $\cal X$ over it,
and calculated the genus $g$ of the curve and the degree of the bundle
(i.e.\ the divisor $\cal D$ degree). As a helpful tool, a matrix $\cal A$
of special simple form~(\ref{119}) was used which, from many viewpoints,
was ``generic enough''. In Section~\ref{1secevd} we will study how
these objects behave under evolution introduced in Section~\ref{1secdef}.

\section{Evolution in terms of divisors}
\label{1secevd}

In this section it is shown, at first, that there exists a one-to-one
correspondence
(more precisely, a birational isomorphism) between the set of block
matrices $\cal A$ (\ref{101}) taken up to gauge transformations (\ref{104}),
and the set of pairs (an algebraic curve, a linear equivalence class of
divisors on it) of a certain kind. This correspondence has, in essence,
been constructed in Section~\ref{1seccrv}, and here are some missing details.
Then, it is explained which divisors and why correspond to the factors
${\cal A}_1, {\cal A}_2$ and
${\cal A}_3 $ in (\ref{102}) taken separately. Finally, it is demonstrated
that to the matrix $\cal B$ \/ (\ref{109}) obtained from $\cal A$
by reversing the order of its factors, the same curve $\Gamma$ corresponds,
but the divisor undergoes some constant shift. Thus, the motion linearizes
in the Jacobian of curve $\Gamma$. Let us proceed to a detailed
consideration.

Equations~(\ref{111}, \ref{112}) define, for a block matrix
$\cal A$, an algebraic curve $\Gamma$. Those equations can obviously be
written as
\begin{eqnarray} \sum^{m}_{i=0} \sum^{n}_{j=0} \sum^{r}_{k=0} a_{ijk}
u^i v^j w^k =0, \vspace{-1.8ex} \nonumber \\
\vspace{-2.8ex} \label{127} \\
v=uw. \nonumber \phantom{u^i v^j w^k =0,} \end{eqnarray}
Besides, a linear bundle over $\Gamma$ has been constructed in
Section~\ref{1seccrv}---the bundle of vectors
$\bmat{c} X\\Y\\Z \emat$
(\ref{113}).
That means that the divisor
$\cal D$ of the bundle is determined, up to linear equivalence, whose
degree is $g+m+n+r-1$, \ $g$ being the curve's genus (\ref{126}).
Gauge transformations (\ref{104}) do not change a pair
($\Gamma$, class of divisor  $\cal D$~).

Now let us show how to construct the matrix $\cal A$ starting from
coefficients
$a_{ijk}$ of the curve (\ref{127}) (arbitrary complex numbers in
``general position'') and a divisor $\cal D$  of degree $g+m+n+r-1$. Note
that
genus $g$ of the curve $\Gamma$ defined by formulae (\ref{127}) without any
(a priori)
connection with block matrices is given by the same formula~(\ref{125}).
This can be seen, e.g., by starting again from the ``simple'' curve
(\ref{120}) of Section~\ref{1seccrv} whose genus is known.
{\em Define} now the meromorphic column vectors
$X,Y$ and $Z$,
guided by Lemma~\ref{1lem3}: for components of vector $X$,
take $m$ linearly independent meromorphic functions on $\Gamma$
satisfying relation~(\ref{116}), and for $Y$ and $Z$
take, similarly, $n$ functions satisfying (\ref{117}) and $r$  functions
satisfying (\ref{118}).

Note also that Lemma~\ref{1lem2} about divisors $(u)_{\infty}, (v)_{\infty},
(w)_{\infty}$ entering in formulae (\ref{116}--\ref{118}) remains valid
for curves defined by an
``abstract'' system (\ref{127}), which is immediately seen on
substituting zero or infinity for $u, v$, or $w$ in
(\ref{127}).

It is clear now that relation (\ref{113})   {\em determines}
unambiguously the matrix $\cal A$ (cf.\ a similar construction in
paper~\cite{Krichever}).
Another choice of linearly independent functions for components of
$X,Y$ and $Z$ leads, of course, to a gauge transformation~(\ref{104}).

Examine now each multiplier in factorization~(\ref{102}) separately.
Lemma~\ref{1lem1}  shows that factorization~(\ref{102}), if exists, is
unique to within the transformations~(\ref{104}).
Let us demonstrate how to construct this factorization by
algebro-geometrical means.

Consider the following figure (Fig.~\ref{1figdecom}).

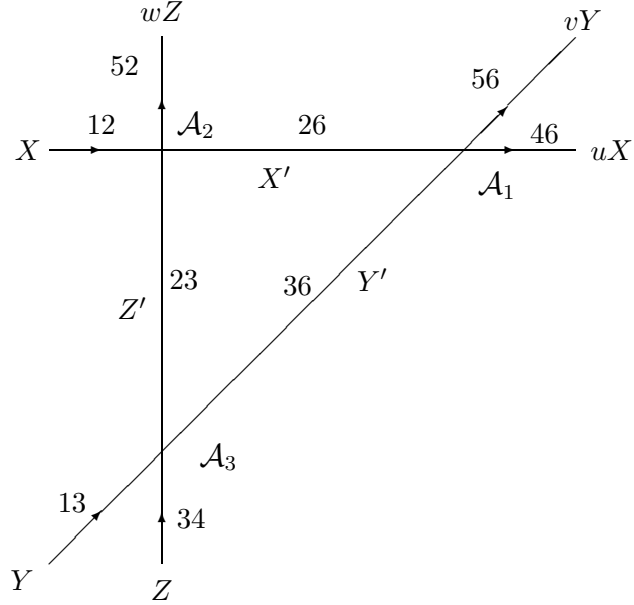
\begin{figure}
\begin{center}
\unitlength=1.00mm
\special{em:linewidth 0.4pt}
\linethickness{0.4pt}
\begin{picture}(76.00,77.42)
\put(19.00,5.17){\vector(0,1){6.88}}
\put(19.00,60.22){\line(0,0){0.00}}
\put(19.00,60.22){\line(0,0){0.00}}
\put(19.00,60.22){\line(0,0){0.00}}
\put(19.00,60.22){\line(0,0){0.00}}
\put(19.00,60.22){\line(0,0){0.00}}
\put(19.00,12.05){\vector(0,1){55.05}}
\put(19.00,67.10){\line(0,1){8.17}}
\put(4.00,60.22){\vector(1,0){7.00}}
\put(11.00,60.22){\vector(1,0){55.00}}
\put(66.00,60.22){\line(1,0){8.00}}
\put(4.00,5.17){\vector(1,1){7.00}}
\put(11.00,12.05){\vector(1,1){54.00}}
\put(65.00,66.24){\line(1,1){9.00}}
\put(3.00,60.22){\makebox(0,0)[rc]{$X$}}
\put(11.00,62.37){\makebox(0,0)[cb]{12}}
\put(16.00,71.40){\makebox(0,0)[rc]{52}}
\put(19.00,77.42){\makebox(0,0)[cb]{$wZ$}}
\put(21.00,62.37){\makebox(0,0)[lb]{\large ${\cal A}_2$}}
\put(39.00,62.37){\makebox(0,0)[cb]{26}}
\put(64.00,68.39){\makebox(0,0)[rb]{56}}
\put(75.00,76.13){\makebox(0,0)[cb]{$vY$}}
\put(68.00,61.08){\makebox(0,0)[lb]{46}}
\put(76.00,60.22){\makebox(0,0)[lc]{$uX$}}
\put(61.00,57.21){\makebox(0,0)[lt]{\large${\cal A}_1$}}
\put(34.00,58.07){\makebox(0,0)[ct]{$X'$}}
\put(39.00,42.16){\makebox(0,0)[rc]{36}}
\put(45.00,43.00){\makebox(0,0)[lc]{$Y'$}}
\put(24.00,19.36){\makebox(0,0)[lc]{\large ${\cal A}_3$}}
\put(21.00,11.19){\makebox(0,0)[lc]{34}}
\put(19.00,3.02){\makebox(0,0)[ct]{$Z$}}
\put(9.00,12.05){\makebox(0,0)[rb]{13}}
\put(2.00,3.02){\makebox(0,0)[rc]{$Y$}}
\put(17.00,39.14){\makebox(0,0)[rc]{$Z'$}}
\put(20.00,43.02){\makebox(0,0)[lc]{23}}
\end{picture}

\end{center}
\caption{Factorization of matrix $\cal A$ and the divisors}
\label{1figdecom}
\end{figure}

The meaning of the numbers standing near the edges in this figure is
as follows: if those numbers are
$jk$, then the meromorphic vector corresponding to the edge consists of such
functions $f$ whose zero and pole divisor $(f)$ satisfies inequality
$$(f)+{\cal D} - {\cal D}_j -{\cal D}_k \geq 0.$$
Those inequalities must be in agreement with
Lemmas~\ref{1lem2} and \ref{1lem3}.
In particular, the matrix ${\cal A}_3$ will be {\em defined}\/ by equality
(notations of formulae~(\ref{103}) are used),
\be \bmat{cc}A_3&B_3\\C_3&D_3\emat \bmat{c}Y\\Z  \emat=
\bmat{c} Y'\\Z' \emat, \label{128} \ee
where the meromorphic vector  $Y$ consists of functions $f$ such that
$$(f)+{\cal D} - {\cal D}_1 - {\cal D}_3 \geq 0$$
(formulae (\ref{117}) and (\ref{114})); $Z$ of functions such that
$$(f)+{\cal D} - {\cal D}_3 - {\cal D}_4 \geq 0$$
(formulae (\ref{118}) and (\ref{114})); $Y'$ and $Z'$  consist
{\em by definition}
of such linearly independent functions that
$$(f)+{\cal D} - {\cal D}_3 - {\cal D}_6 \geq 0$$
for $Y'$, and
$$(f)+{\cal D} - {\cal D}_2 - {\cal D}_3 \geq 0$$
for $Z'$. It is easy to see that (\ref{128})  is a correct definition
for matrix ${\cal A}_3$, because the components of each of the vectors
$\bmat{c}Y\\Z \emat$ and $\bmat{c}Y'\\Z' \emat$ form a basis in the space of
meromorphic functions $f$ such that
$$(f)+{\cal D} - {\cal D}_3  \geq 0.$$

Next, let
\be  \bmat{cc} A_2&B_2\\ C_2& {\cal D}_2 \emat
\bmat{c} X\\Z' \emat = \bmat{c} X'\\wZ \emat, \label{129}  \ee
\be  \bmat{cc} A_1&B_1\\ C_1& {\cal D}_1 \emat
\bmat{c} X'\\Y' \emat = \bmat{c}uX\\vY \emat, \label{130} \ee
where $X'$ consists of functions $f$ such that
$$(f)+{\cal D} - {\cal D}_2 - {\cal D}_6 \geq 0.$$
It is shown in much the same way as above that equalities~(\ref{129}) and
(\ref{130}) do correctly define the matrices ${\cal A}_2$ and ${\cal A}_1$.
What remains is to check the validity of equality~(\ref{102}) for
${\cal A}_1$, ${\cal A}_2$ and ${\cal A}_3$ given by these definitions.
To do this, observe that (\ref{128}--\ref{130}) together yield
\be {\cal A}_1{\cal A}_2{\cal A}_3 \bmat{c} X\\Y\\Z \emat=
\bmat{c} uX\\vY\\wZ \emat . \label{131} \ee
The equality~(\ref{102}) follows from comparing~(\ref{131}) with (\ref{113}).

Note that the arbitrariness in choosing $X',Y'$ and $Z'$ corresponds,
of course, to transformations~(\ref{104}).

Now let us pass to matrix $\cal B$, a product of the same three factors
in the inverse order. The formulae (\ref{109}) and (\ref{128}--\ref{130})
together yield (if one multiplies both sides of (\ref{129}) by $u$, and
both sides of (\ref{128}) by $v$~):
\be {\cal B} \bmat{c} X'\\Y'\\uZ' \emat =
\bmat{c} uX'\\vY'\\vZ' \emat . \label{132} \ee
Compare the divisors of meromorphic vectors in L.H.S.'s of
(\ref{132}) and (\ref{113}). An easy calculation shows that
$$ {\cal D}_{X'}-{\cal D}_X = {\cal D}_{Y'}-{\cal D}_Y =
{\cal D}_{(uZ')}-{\cal D}_Z = {\cal D}_1 - {\cal D}_6.$$
One sees hence that the same curve $\Gamma$ corresponds to the operator
${\cal B}$ as to the operator ${\cal A}$, while the divisor ${\cal D}$
changes to ${\cal D}+{\cal D}_1-{\cal D}_6 $.

Thus, in this section the name ``invariant'' has been justified for the
curve $\Gamma$: it has been shown not to change under the evolution of
Section~\ref{1secdef}. At the same time, it was demonstrated how to construct
the factorization~(\ref{102}). Finally, it was shown that
the evolution is described in algebro-geometrical terms
 as a linear, with respect to discrete time,
change of (the linear equivalence class of) divisor ${\cal D}$: it changes
by ${\cal D}_1-{\cal D}_6$ per each unit of time.

\begin{sloppypar}
\section{Connection with the 6-vertex model on the kagome lattice}
\end{sloppypar}
\label{1seckag}

\begin{figure}
\begin{center}
\unitlength=0.50mm
\special{em:linewidth 0.4pt}
\linethickness{0.4pt}
\begin{picture}(91.00,91.99)
\put(16.00,15.91){\circle{4.00}}
\put(46.00,15.91){\circle{4.00}}
\put(76.00,15.91){\circle{4.00}}
\put(76.00,46.02){\circle{4.00}}
\put(46.00,46.02){\circle{4.00}}
\put(16.00,46.02){\circle{4.00}}
\put(16.00,76.13){\circle{4.00}}
\put(46.00,76.13){\circle{4.00}}
\put(76.00,76.13){\circle{4.00}}
\put(91.00,76.13){\line(-1,0){13.00}}
\put(74.00,76.13){\line(-1,0){26.00}}
\put(44.00,76.13){\line(-1,0){26.00}}
\put(14.00,76.13){\line(-1,0){13.00}}
\put(1.00,46.02){\line(1,0){13.00}}
\put(18.00,46.02){\line(1,0){26.00}}
\put(48.00,46.02){\line(1,0){26.00}}
\put(78.00,46.02){\line(1,0){13.00}}
\put(91.00,15.91){\line(-1,0){13.00}}
\put(74.00,15.91){\line(-1,0){26.00}}
\put(44.00,15.91){\line(-1,0){26.00}}
\put(14.00,15.91){\line(-1,0){13.00}}
\put(16.00,1.29){\line(0,1){12.04}}
\put(16.00,13.33){\line(0,1){0.86}}
\put(16.00,18.07){\line(0,1){26.24}}
\put(16.00,48.17){\line(0,1){25.81}}
\put(16.00,78.28){\line(0,1){12.90}}
\put(46.00,91.18){\line(0,-1){12.90}}
\put(46.00,73.98){\line(0,-1){25.81}}
\put(46.00,44.30){\line(0,-1){26.24}}
\put(46.00,14.19){\line(0,-1){12.90}}
\put(76.00,1.29){\line(0,1){12.90}}
\put(76.00,18.07){\line(0,1){26.24}}
\put(76.00,48.17){\line(0,1){25.81}}
\put(76.00,78.28){\line(0,1){12.90}}
\put(18.00,47.31){\line(1,1){27.00}}
\put(48.00,76.99){\line(1,1){15.00}}
\put(48.00,47.31){\line(1,1){27.00}}
\put(78.00,76.99){\line(1,1){13.00}}
\put(78.00,47.31){\line(1,1){13.00}}
\put(78.00,17.21){\line(1,1){13.00}}
\put(1.00,1.29){\line(1,1){13.00}}
\put(31.00,1.29){\line(1,1){13.00}}
\put(61.00,1.29){\line(1,1){13.00}}
\put(1.00,30.97){\line(1,1){13.00}}
\put(13.00,43.01){\line(4,5){1.00}}
\put(1.00,61.08){\line(1,1){13.00}}
\put(31.00,91.18){\line(-1,-1){13.00}}
\put(18.00,18.06){\line(1,1){26.00}}
\put(48.00,18.06){\line(1,1){26.00}}
\end{picture}

\end{center}
\caption{The triangular lattice}
\label{1figtria}
\end{figure}
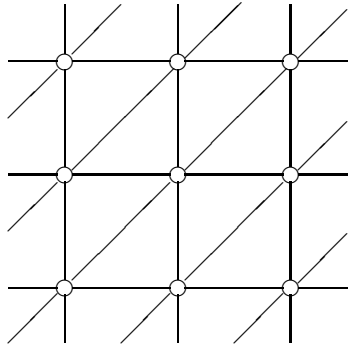

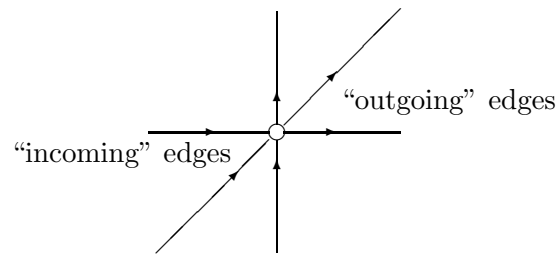
\begin{figure}
\begin{center}
\unitlength=0.50mm
\special{em:linewidth 0.4pt}
\linethickness{0.4pt}
\begin{picture}(67.00,65.61)
\put(34.00,32.70){\circle{4.00}}
\put(34.00,12.48){\vector(0,1){13.33}}
\put(34.00,25.82){\line(0,1){4.73}}
\put(34.00,34.85){\vector(0,1){9.03}}
\put(34.00,43.88){\line(0,1){10.75}}
\put(51.00,40.87){\makebox(0,0)[lc]{``outgoing'' edges}}
\put(22.00,26.68){\makebox(0,0)[rc]{``incoming'' edges}}
\put(24.00,22.81){\line(1,1){8.00}}
\put(13.00,11.62){\vector(1,1){11.00}}
\put(0.00,32.69){\vector(1,0){18.00}}
\put(18.00,32.69){\line(1,0){14.00}}
\put(36.00,32.69){\vector(1,0){14.00}}
\put(50.00,32.69){\line(1,0){17.00}}
\put(34.00,0.86){\line(0,1){12.04}}
\put(34.00,54.63){\line(0,1){10.32}}
\put(36.00,34.84){\vector(1,1){14.00}}
\put(50.00,48.61){\line(1,1){17.00}}
\put(13.00,11.61){\line(-1,-1){11.00}}
\end{picture}

\end{center}
\caption{The vectors corresponding to ``incoming'' edges are transformed by
$\cal A$ into linear combinations of those corresponding to ``outgoing''
edges}
\label{1figinout}
\end{figure}

Consider now a reduction of the system defined in Section~\ref{1secdef}
which leads to a dynamical system in  $2+1$-dimensional fully discrete
space-time. Let the linear space in which matrix $\cal A$~(\ref{101})
acts have a basis enumerated by edges of a triangular lattice
on the torus (Fig.~\ref{1figtria}).
Let the lattice contain $m$ horisontal edges, $n$ oblique edges and $r$
vertical edges, and let this correspond to the block structure (\ref{101})
of matrix $\cal A$ in the sense that, say, the block $A$ acts in the space
generated by basis vectors corresponding to horisontal edges and so on.
Impose the following ``locality'' condition on matrix $\cal A $:
let the vector corresponding  to any given edge of the lattice be
transformed under the action of  $\cal A$ into a linear combination of
just three vectors, corresponding to the edges coming upwards, to the right
and  northeastwards from the vertex that is the upper, right, or
northeastern end of the considered ``incoming'' edge (Fig.~\ref{1figinout}).
Thus, only those elements of matrix $\cal A$ are not zeros that correspond to
``local'' transitions of Fig.~\ref{1figinout}.

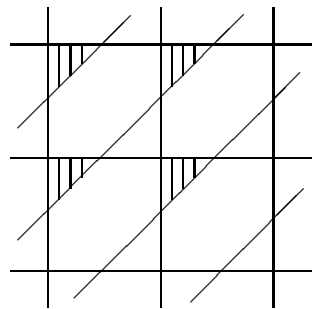
\begin{figure}
\begin{center}
\unitlength=0.50mm
\special{em:linewidth 0.4pt}
\linethickness{0.4pt}
\begin{picture}(80.00,80.86)
\put(10.00,0.86){\line(0,1){40.00}}
\put(10.00,40.86){\line(0,0){0.00}}
\put(10.00,40.86){\line(0,0){0.00}}
\put(10.00,40.86){\line(0,0){0.00}}
\put(10.00,40.86){\line(0,0){0.00}}
\put(10.00,40.86){\line(0,0){0.00}}
\put(10.00,40.86){\line(0,0){0.00}}
\put(10.00,40.86){\line(0,0){0.00}}
\put(10.00,40.86){\line(0,0){0.00}}
\put(10.00,40.86){\line(0,0){0.00}}
\put(10.00,40.86){\line(0,0){0.00}}
\put(10.00,40.86){\line(0,0){0.00}}
\put(10.00,40.86){\line(0,0){0.00}}
\put(10.00,40.86){\line(0,0){0.00}}
\put(10.00,40.86){\line(0,0){0.00}}
\put(10.00,40.86){\line(0,0){0.00}}
\put(10.00,40.86){\line(0,0){0.00}}
\put(10.00,40.86){\line(0,1){40.00}}
\put(40.00,80.86){\line(0,-1){80.00}}
\put(70.00,0.86){\line(0,1){80.00}}
\put(0.00,10.75){\line(1,0){80.00}}
\put(80.00,40.86){\line(-1,0){80.00}}
\put(0.00,70.97){\line(1,0){80.00}}
\put(2.00,48.60){\line(1,1){30.00}}
\put(2.00,18.93){\line(1,1){60.00}}
\put(17.00,3.44){\line(1,1){60.00}}
\put(48.00,2.58){\line(1,1){30.00}}
\put(13.00,70.97){\line(0,-1){11.18}}
\put(16.00,62.80){\line(0,1){8.17}}
\put(19.00,70.97){\line(0,-1){5.16}}
\put(13.00,40.86){\line(0,-1){11.18}}
\put(16.00,32.69){\line(0,1){8.17}}
\put(19.00,40.86){\line(0,-1){5.16}}
\put(43.00,40.86){\line(0,-1){11.18}}
\put(46.00,32.69){\line(0,1){8.17}}
\put(49.00,40.86){\line(0,-1){5.16}}
\put(43.00,70.97){\line(0,-1){11.18}}
\put(46.00,62.80){\line(0,1){8.17}}
\put(49.00,70.97){\line(0,-1){5.16}}
\end{picture}

\end{center}
\caption{The kagome lattice}
\label{1figkagome}
\end{figure}

The factorization of a ``local'' matrix  $\cal A$ into the product
(\ref{102}) corresponds to each vertex represented by a small circle in
Fig.~\ref{1figtria} being converted into a triangle of the type shown in
Fig.\ref{1figdecom}, so that the lattice transforms into a kagome lattice
(Fig.~\ref{1figkagome}). The triangles arising from the vertices-circles
are shaded in Fig.~\ref{1figkagome}.

A step of the evolution consists in those triangles being ``turned
inside out'', or else one may say that each oblique line
replaces the one above it. We will see soon that this procedure is connected
with solutions of the ``local'' in the sense of \cite{MN} quantum
Yang---Baxter equation.

Now return to the triangle lattice of Figure~\ref{1figtria}.
Let us express the ``integral of motion''
\be I(u,w)= \det \bigl({\bf 1}-{\cal A} \bmat{ccc} u^{-1}&0&0\\
0&u^{-1}w^{-1}&0 \\ 0&0&w^{-1} \emat \bigr) \label{133} \ee
in terms of paths going along the edges of this lattice ($I(u,w)$
is indeed an integral of motion with any $u,w$, because the equality
$I(u,w)=0$ determines the invariant curve, and a possible
multiplicative constant is fixed by the fact that the constant term
in (\ref{133}) equals unity).

As it known, the determinant of a martix is an alternating sum of
its elements' products, each summand corresponding to some permutation
of the matrix columns, while each permutation  factorizes into a product
of cyclic ones. As applied to our matrix $\cal A$, it means that the
determinant~(\ref{133}) is a sum each term of which corresponds to a set
of closed trajectories going along the arrows according to
Fig.~\ref{1figinout} (recall that the lattice is situated on the torus!).
The trajectories of each given set can have intersections and
self-intersections, but none of the {\em edges} may be passed through twice
or more by one or several trajectories.

To be exact, to each trajectory corresponds a product of entries of
the matrix
$${\cal A} \bmat{ccc} u^{-1}&0&\\ 0&u^{-1}w^{-1}&0\\
0&0&w^{-1} \emat $$
corresponding to transitions through a vertex to a neighboring edge
according to Fig.~\ref{1figinout}, multiplied (the product as a whole)
by  $(-1)$. To each set of trajectories (including, of course, the empty
set) corresponds the product of the mentioned values corresponding to
its trajectories. The reader can verify that all the minus signs,
including that in formula~(\ref{133}), are taken into account correctly.

It is easy also to describe the determinant $I(u,w)$ in terms of the kagome
lattice obtained on factorizing the matrix  $\cal A$ into the
product~(\ref{102}).
This description almost repeats two preceding paragraphs. Let us formulate it
as the following lemma.
\begin{lemma}
\label{1lemdet}
$I(u,w)$ is a sum over sets of trajectories on the kagome lattice; the
direction of motion is upwards,  to the right, or northeastwards;
none of the edges is passed through twice by trajectories of a given set;
to the vertices of types
\unitlength=0.4mm
\special{em:linewidth 0.4pt}
\linethickness{0.4pt}
\begin{picture}(15.00,9.46)
\put(1.00,4.73){\line(1,0){14.00}}
\put(1.00,0.00){\line(3,2){14.00}}
\end{picture},
\unitlength=0.4mm
\special{em:linewidth 0.4pt}
\linethickness{0.4pt}
\begin{picture}(14.00,12.04)
\put(0.00,6.02){\line(1,0){14.00}}
\put(7.00,0.00){\line(0,1){12.04}}
\end{picture},
\unitlength=0.4mm
\special{em:linewidth 0.4pt}
\linethickness{0.4pt}
\begin{picture}(9.00,11.18)
\put(4.00,0.00){\line(0,1){11.18}}
\put(0.00,0.00){\line(4,5){9.00}}
\end{picture}
\/, if a trajectory passes through them, correspond the factors
equalling matrix elements of matrices  ${\cal A}_1$, ${\cal A}_2$,
${\cal A}_3$ respectively; besides, to each move to the right through
a lattice period corresponds a factor $u^{-1}$, and to each move
upwards---a factor $w^{-1}$ (and both of them to a diagonal move);
finally, to each  set corresponds one more factor,
$(-1)^{\mbox{\small(number of trajectories)}}$ .
\end{lemma}

Now let us link  $I(u,w)$ with the statistical sum of (inhomogeneous)
6-vertex model on the kagome lattice. Let each edge of the kagome lattice
be able to take one of two states, which will be depicted below
as either presence or absence of an arrow on the edge (the arrow will
always be directed upwards, to the right, or northeastwards).
A ``Boltzmann weight'' will correspond to each vertex as follows:
if there are no arrows on the edges meeting at the vertex,
the weight will be $1$; if there is exactly one arrow coming into
the vertex and exactly one going out of it, the weight will be equal to
the corresponding matrix element of  ${\cal A}_1$, ${\cal A}_2$
or ${\cal A}_3$ (e.g.\, to the vertex
\unitlength=0.60mm
\special{em:linewidth 0.4pt}
\linethickness{0.4pt}
\begin{picture}(20.00,12.00)
\put(0.00,2.00){\vector(1,0){5.00}}
\put(5.00,2.00){\line(1,0){15.00}}
\put(10.00,-8.00){\vector(0,1){15.00}}
\put(10.00,7.00){\line(0,1){5.00}}
\end{picture}
\vspace{3mm}
\ \ corresponds the weight equal to the matrix element of ${\cal A}_2$
that is responsible for a transition between the vector
corresponding to the left edge, and the vector correcponding to the upper
edge); if there are 2 incoming and 2 outgoing arrows, the weight is
the difference between the products of weights corresponding to the
intersecting and non-intersecting paths through the vertex:
\be \ba{c} \mbox{Weight} \left(
\unitlength=0.60mm
\special{em:linewidth 0.4pt}
\linethickness{0.4pt}
\begin{picture}(20.00,12.00)
\put(0.00,2.00){\vector(1,0){6.00}}
\put(6.00,2.00){\vector(1,0){10.00}}
\put(6.00,2.00){\line(1,0){14.00}}
\put(10.00,-8.00){\vector(0,1){6.00}}
\put(10.00,-2.00){\vector(0,1){10.00}}
\put(10.00,8.00){\line(0,1){4.00}}
\end{picture}
\right)=\mbox{Weight}
\left(
\unitlength=0.60mm
\special{em:linewidth 0.4pt}
\linethickness{0.4pt}
\begin{picture}(20.00,12.00)
\put(0.00,2.00){\vector(1,0){5.00}}
\put(5.00,2.00){\line(1,0){15.00}}
\put(10.00,-8.00){\vector(0,1){15.00}}
\put(10.00,7.00){\line(0,1){5.00}}
\end{picture}
 \right)
\cdot \mbox{Weight}\left(
\unitlength=0.60mm
\special{em:linewidth 0.4pt}
\linethickness{0.4pt}
\begin{picture}(20.00,12.00)
\put(0.00,2.00){\vector(1,0){15.00}}
\put(15.00,2.00){\line(1,0){5.00}}
\put(10.00,-8.00){\vector(0,1){5.00}}
\put(10.00,-3.00){\line(0,1){15.00}}
\end{picture}
 \right)- \\
-\mbox{Weight} \left(
\unitlength=0.60mm
\special{em:linewidth 0.4pt}
\linethickness{0.4pt}
\begin{picture}(20.00,12.00)
\put(0.00,2.00){\vector(1,0){6.00}}
\put(6.00,2.00){\vector(1,0){10.00}}
\put(16.00,2.00){\line(1,0){4.00}}
\put(10.00,-8.00){\line(0,1){20.00}}
\end{picture}
 \right) \cdot
\mbox{Weight}\left(
\unitlength=0.60mm
\special{em:linewidth 0.4pt}
\linethickness{0.4pt}
\begin{picture}(20.00,12.00)
\put(0.00,2.00){\line(1,0){20.00}}
\put(10.00,-2.00){\vector(0,1){10.00}}
\put(10.00,8.00){\line(0,1){4.00}}
\put(10.00,-8.00){\vector(0,1){6.00}}
\end{picture}
 \right); \ea
\label{134} \ee
in the rest of cases the weight is zero.

A weight will also correspond to each edge of the kagome lattice:
weight $1$ to an edge without an arrow, and weights  $u^{-1/2}$,
$w^{-1/2}$ or  $u^{-1/2}w^{-1/2}$ to a horizontal, vertical or oblique
edge having an arrow. If needed, the edge weights can be included in the
vertex weights, but we will not do that here.

The statistical sum $S(u,w)$ of our 6-vertex model is, of course, a sum
of products of vertex and edge weights over all arrow cinfigurations.
The next lemma is the key statement.

\begin{lemma} \label{1lemsta}
The statistical sum $S(u,w)$ is a sum over the same sets of trajectories as
the determinant $I(u,w)$, and to each set corresponds the same summand
up to, maybe, a minus sign. To be exact, the {\rm number of trajectories}
 in the
exponent of $(-1)$ in Lemma~\ref{1lemdet}
changes to the {\rm number of intersections} (self-intersections included)
of a given set of trajectories.
\end{lemma}

{\it Proof}\/ is evident from the statistical sum definition.

Each closed path on the torus is homologically equivalent to a linear
combination of two basis cycles ${\bf a}$ and ${\bf b}$. The same is true
for a set of paths (trajectories), regarded as a formal sum of them.
Different sets may be homologically equivalent to a given cycle
$l{\bf a}+m{\bf b}$ but, as the following lemma shows, they have something
in common.
\begin{lemma} \label{1lemtor}
For any set of trajectories on the torus homologically equivalent to a cycle
$l{\bf a}+m{\bf b}$ (${\bf a},{\bf b}$ being basis cycles, $l,m$---integers),
\be (\mbox{\rm number of intersections})-(\mbox{\rm number of trajectories})
\equiv lm-l-m(\mbox{mod} 2).
\label{135} \ee
\end{lemma}

{\it Proof}\/ may consist in the following simple consideration:
1)~if the set consists of $l$ trajectories going along ${\bf a}$, and $m$
ones going along ${\bf b}$, (\ref{135}) is obviously true,
2) under deformations of trajectories, the number
of intersections changes only by even numbers,
3) with elimination of an intersection $\left(
\unitlength=0.50mm
\special{em:linewidth 0.4pt}
\linethickness{0.4pt}
\begin{picture}(69.00,12.00)
\put(10.00,-8.00){\line(0,1){20.00}}
\put(0.00,2.00){\line(1,0){20.00}}
\put(24.00,2.00){\vector(1,0){10.00}}
\put(39.00,12.00){\oval(20.00,20.00)[rb]}
\put(59.00,-8.00){\oval(20.00,20.00)[lt]}
\end{picture}
\right)$
or inverse operation, the L.H.S. of (\ref{135}) may change also only by an
even number. Starting from an arbitrary set and applying the
transformations 2) and 3), one can arrive at a set of type 1),
so the lemma is proved.

Let the following products of edge weights correspond to the basis cycles:
$x=u^{\alpha_1}w^{\beta_1}$  for ${\bf a}$ and
$y=u^{\alpha_2}w^{\beta_2}$ for ${\bf b}$. Denote
\be s(x,y)=S(u,w), \quad f(x,y)=I(u,w). \label{136} \ee

\begin{theorem}
\label{1thstadet}
The statistical sum of the inhomogeneous 6-vertex model on the kagome lattice
defined in this section is invariant with respect to the evolution of the
reduced
$2+1$-dimensional model (for all $u,w$) and is connected with the
determinant $I(u,w)$~(\ref{133}), whose vanishing defines the invariant curve
of the model, by relations (in the notations of (\ref{136}))
\be s(x,y)=1/2 \left( -f(x,y)+f(-x,y)+f(x,-y)+f(-x,-y) \right), \label{137}
\ee
\be f(x,y)=1/2 \left( -s(x,y)+s(-x,y)+s(x,-y)+s(-x,-y) \right). \label{138}
\ee
\end{theorem}

{\it Proof.} It follows from Lemmas~\ref{1lemdet}, \ref{1lemsta},
\ref{1lemtor} that in the expansions of $s(x,y)$ and  $f(x,y)$ in powers of
$x$ and $y$ the coefficients  near $x^l y^m$ coincide if $lm-l-m$ is even,
and differ in their signs in the opposite case. This is exactly what the
formulae~(\ref{137},\ref{138}) are about.

\section{Discussion}

The reader can verify that the 6-vertex model weight matrices corresponding
to vertices of a triangle of the kagome lattice, and those matrices
corresponding to the same triangle ``turned inside out'', indeed satisfy
a local Yang---Baxter equation of the type of~(\ref{1lqyb}). Such local, or
generalized\footnote{also called gauged or modified}, equations are gaining
popularity, and now not only solutions to the generalized Yang---Baxter
equation, but even those for generalized tetrahedron equation have been
found \cite{KashStrog,MS,MSS}. The author hopes to show soon that the
constructions in this paper do have relationship with the tetrahedron
equation and its generalizations.

It is interesting also to link our model on the kagome lattice with the
model generated by the star---triangle transformation~\cite{Baxterbook} and
described in the end of the paper~\cite{Korepanov-dimers}. It turns out
that, for that purpose, one has at first to examine the case of {\em
orthogonal\/} matrix $\cal A$~(\ref{101},\ref{102}) (its factors
${\cal A}_1,{\cal A}_2,{\cal A}_3$ being orthogonal, too). The orthogonality
corresponds to interesting symmetry conditions imposed on the invariant
curve~$\Gamma$ and divisor~$\cal D$\/: the {\em canonical class} of $\Gamma$
is needed to formulate the condition on $\cal D$ in the most elegant way!
The details about the star---triangle model, orthogonality etc.\ will be
presented in one of the author's coming publications.

Thus, it was a certain generalization of the star---triangle transformation
that has been presented in this paper, and one more aspect of close
connection between the classical and quantum integrable models has been
demonstrated.

\end{document}